\documentclass[aps,pra,twocolumn,showkeys,nofootinbib,superscriptaddress]{revtex4}
\pdfoutput=1

\usepackage[utf8]{inputenc}
\usepackage{graphicx}
\usepackage{graphics}
\usepackage{epsfig}
\usepackage{lscape}
\usepackage[dvipsnames]{xcolor}
\usepackage{color}
\usepackage{multirow}
\usepackage{hyperref}
\usepackage{breqn}
\usepackage{physics}
\usepackage{subfiles}
\usepackage{ragged2e}
\usepackage{amssymb}
\usepackage{multirow}
\usepackage{comment}
\usepackage{balance}

\newcommand{\Chemie}{
Johannes Gutenberg-Universit\"at, Fritz-Strassmann Weg 2, 55128 Mainz, Germany \\
}

\newcommand{\HIM}{
Helmholtz-Institut Mainz,
Staudingerweg 18, 55128 Mainz, Germany \\
}

\newcommand{\GSI}{
GSI Helmholtzzentrum f\"ur Schwerionenforschung,
Planckstrasse 1, 64291 Darmstadt, Germany\\
}

\newcommand{\GANIL}{
GANIL, CEA/DRF-CNRS/IN2P3, 
B.P. 55027, 14076 Caen Cedex 05, France\\
}

\newcommand{\NBI}{
Niels Bohr Institute, University of Copenhagen,
Rådmandsgade 64, 2200 Copenhagen N, Denmark\\
}

\begin{document}
\title{Electronic State Chromatography of Lutetium Cations}

\author{Biswajit Jana} \thanks{Corresponding author: bjana@uni-mainz.de}
\affiliation{\Chemie} \affiliation{\HIM} \affiliation{\GSI}

\author{EunKang Kim}
\affiliation{\Chemie} \affiliation{\HIM} \affiliation{\GSI}

\author{Aayush Arya}
\affiliation{\Chemie} \affiliation{\HIM} \affiliation{\GSI} \affiliation{\NBI}

\author{Michael Block}
\affiliation{\Chemie} \affiliation{\HIM} \affiliation{\GSI} 

\author{Sebastian Raeder}
\affiliation{\HIM} \affiliation{\GSI} 

\author{Harry Ramanantoanina}
\affiliation{\Chemie} \affiliation{\HIM} \affiliation{\GSI}

\author{Elisabeth Rickert}
\affiliation{\Chemie} \affiliation{\HIM} \affiliation{\GSI}

\author{Elisa Romero Romero}
\affiliation{\Chemie} \affiliation{\HIM} \affiliation{\GSI}

\author{Mustapha Laatiaoui} 
\affiliation{\Chemie} \affiliation{\HIM} \affiliation{\GSI} \affiliation{\GANIL}

\date{\today}

\begin{abstract}
Relativistic effects have a profound impact on the electronic structure of the heaviest elements, and consequently on their chemical and physical properties. In order to investigate these effects, we developed an ion mobility spectrometer equipped with a cryogenic drift tube to measure the reduced mobilities of the ions of monoatomic lanthanides and actinides. As a benchmark, we conducted the first systematic study of lutetium cations (Lu$^{+}$) drifting in helium at $298$K, resolving ground and metastable electronic states using electronic state chromatography. We compared the state-specific mobilities with predictions based on state-of-the-art \textit{ab initio} calculations, and they show very good agreement.
Our work paves the way for atomic structure studies in the region of the superheavy elements and provides a powerful experimental approach for exploring ion-neutral interactions and electronic configuration effects in heavy element systems.

\end{abstract}

\maketitle

{\addtolength{\textheight}{1\baselineskip} 

\section{Introduction}
\vspace{-5mm} 
The study of the physical and chemical properties of the heaviest elements offers fascinating insights into how these atoms behave and interact~\cite{Schadel:2015, Pyykko:2012}. 
As these atoms exhibit a large nuclear charge, their accurate atomic modeling is hardly conceivable without taking relativistic effects into account~\cite{Tatewaki:2017, Schwerdtfeger:2015}.
With increasing atomic number $Z$, the velocity of the $s$ and $p_{1/2}$ inner shell electrons orbiting the nucleus increases along with their effective mass. To conserve angular momentum, their orbitals contract towards the nucleus~\cite{Das:2023}. This contraction effectively shields the nuclear potential, affecting the binding energy of the electrons, their configuration, interatomic forces, bond lengths, and enthalpies~\cite{Wybourne:2002}. 
Consequently, the distinct chemical and physical properties of these elements can deviate significantly from traditional chemical trends.
This makes it more curious to investigate these effects in the lanthanides and actinides at the bottom of the Periodic Table as a stepping stone towards electronic structure investigations in the region of superheavy elements (SHEs)~\cite{Laatiaoui:2019}.

Gas-phase ion transport has long been considered as a sensitive probe of the underlying electronic configuration~\cite{Kemper:1991, Wilkins:2010} and thus of deviations from expected chemical trends due to these relativistic effects. Consequently, Ion Mobility Spectrometry (IMS) is experiencing a revival in the investigation of the heaviest elements~\cite{Laatiaoui:2012, Backe:2015, Manard:2017, Rickert:2020, Block:2021}, sparked after the successful laser spectroscopy of the chemical element 100, fermium~\cite{Sewtz:2003b} using the Ion Guide Resonance Ionization Spectroscopy (IGRIS) technique in a buffer gas cell. In recent years, the combination of laser spectroscopy with IMS has led to the development of a promising approach known as Laser Resonance Chromatography (LRC)~\cite{Laatiaoui:2020a,Laatiaoui:2020b, Romero:2022,Kim:2024}. 
With this emerging technique, electronic structure investigations could potentially become possible in the future on element 103, lawrencium, and perhaps on SHEs as well.
In this respect, IMS offers undoubtedly many advantages~\cite{Wilkins:2010, Valerie:2021}. Being already a powerful tool in structural analysis, its simplicity and high sensitivity make it very promising for probing the electronic structure of monoatomic ions and their ion-neutral interaction potentials even on the production scale of one atom at a time.
\begin{figure*}[tb!]
\centering
\includegraphics[width=0.95\textwidth] {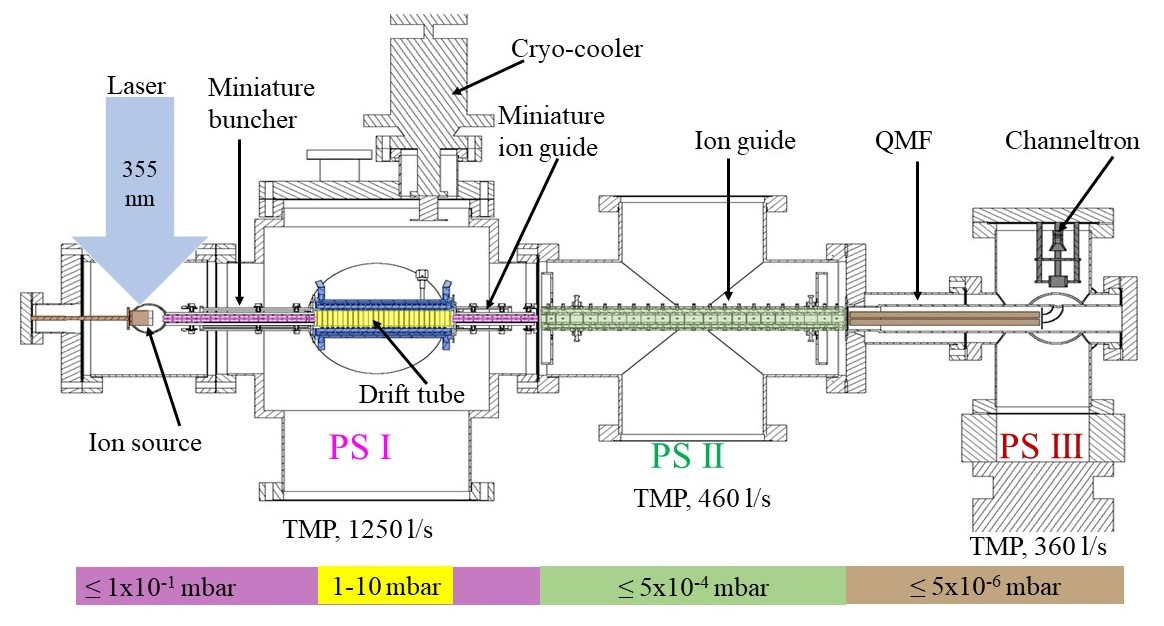}
\caption{Schematic overview of the developed ion mobility spectrometer. Typical pressure ranges are indicated for each pumping section of the apparatus. A third-harmonic laser radiation of a Nd:YAG laser is used during the IMS studies to produce the sample ions from metal foils.}
\label{fig: experimental setup}
\end{figure*}

At its core, ion mobility describes the coefficient that relates the velocity of an ion drifting in a buffer gas to the uniform electric field $E$ that causes the drift. This coefficient is commonly reported as reduced mobility $K_0$, at standard pressure $P_0$ and temperature $T_0$ \cite{Valerie:2021}. From the kinetic theory of gases, the mobility of the ion is sensitive to the momentum transfer collision cross section, which in turn depends on the ion-neutral interaction potential~\cite {Mason:1988}. 
Former IMS studies on lighter elements revealed a deviation in mobility whenever there was a drastic change in electronic configuration. Across lanthanides, for example, the occupation of the $d$ orbital makes the cations faster in helium (He). It leads to a $11$\% deviation in ion mobilities between Eu$^+$ ([Xe]$4f^76s^1$) and Gd$^+$ ([Xe]$4f^75d^16s^1$)~\cite{Manard:2017b}. Even for a single atomic species, different electronic configurations of the ground and low lying metastable states have been identified with different mobility data due to their distinct transport properties~\cite{Kemper:1991, Wilkins:2010}.
This phenomenon, known as the electronic-state chromatography effect, or ESC effect for short~\cite{Kemper:1991}, enables state-selected ion chemistry and has paved the way for studies of state-resolved reaction constants and the electron affinities of reactants~\cite{Bowers:1993,Wilkins:2010,Armentrout:2011}. To enhance the IMS resolution, it is essential to operate the drift tube at cryogenic temperatures~\cite{Wilkins:2010}. The newly developed technique of LRC exploits exactly this ESC effect to explore the electronic structure of the heaviest elemental cations~\cite{Laatiaoui:2020a,Laatiaoui:2020b}. A substantial difference in state-specific ion mobilities for Lr$^+$ ~\cite{Ramanantoanina:2023}and Rf$^+$ ~\cite{Giorgio:2024} cations has been predicted using \textit{ab initio} calculations, illustrating the promise of this technique. 
This new laser spectroscopy technique is still in development, but the initial results of the inauguration experiments are very promising~\cite{Kim:2024}. 

In this paper, we describe a cryogenic drift tube-based ion mobility spectrometer specifically developed to study the transport properties of heavy elemental cations and thereby identify the best candidates for electronic state chromatography. We structured it as follows: The experimental setup is briefly presented in Sec.~\ref{sec:setup}. In Sec.~\ref{sec:results}, we report the chromatography performance of the medium-sized cryogenic drift tube of the setup and systematic ion mobility measurements on Lu$^{+}$ as a case study. In Sec.~\ref{sec:summary}, we conclude with a short summary and outlook.

\section{Experimental Setup}\label{sec:setup}
\vspace{-5mm}
Figure~\ref{fig: experimental setup} shows a schematic of the developed ion mobility spectrometer. It consists of three different pressure sections (PS) interconnected by orifices of $3\,$mm diameter, starting with the chromatography section (PS1) and ending with the ion detection chamber (PS3).
The chromatography section incorporates, in sequence, a laser-ablation ion source, a miniature Radio-Frequency (RF) ion buncher, a cryogenic drift tube, and a miniature RF ion guide. This section is evacuated by a turbo-molecular pump (TMP, type: Edwards STPA-1603C) with a maximum pumping speed of $1250\,$l/s for helium gas. Typical background pressure of this first section is of the order of $10^{-2}$ mbar.
The detection chamber houses a quadrupole mass filter (QMF, type: Balzers QMG 311) and a channeltron detector (type: Sjuts, KBL 15RS), installed at a $90^{\circ}$ angle to the QMF axis. It is evacuated by a TMP (type: Leybold TURBOVAC 361, $360\,$l/s) of a moderate pumping speed. A detailed description of this detection chamber can be found in Refs.~\cite{Backe:1997,Sewtz:2003} as it  was formally part of the IGRIS apparatus used for fermium spectroscopy.
Since the operation of the channeltron requires background pressures on the order of $10^{-6}$ mbar, a second intermediate section (PS2) containing an RF ion guide is used to establish differential pumping between the former sections (PS1 and PS3). This PS2 section is evacuated by another TMP (type: Edwards STP 451, $460\,$l/s). 
A dry pump (Edwards iXL600) with a pumping speed of $600\,$m$^3$/h is used as a backing pump for the entire setup.
The ion mobility spectrometer is designed for mobility measurements of heavy ions in different noble gases at temperatures ranging from $70$ to $400$K and at reduced electric fields $E/n_0\leq 20\,$Td with $1\,$Td (Townsend unit) being equal to $10^{-17}\,$Vcm$^2$. 
\begin{figure}[bt!]
\centering
\includegraphics[width=0.48\textwidth] {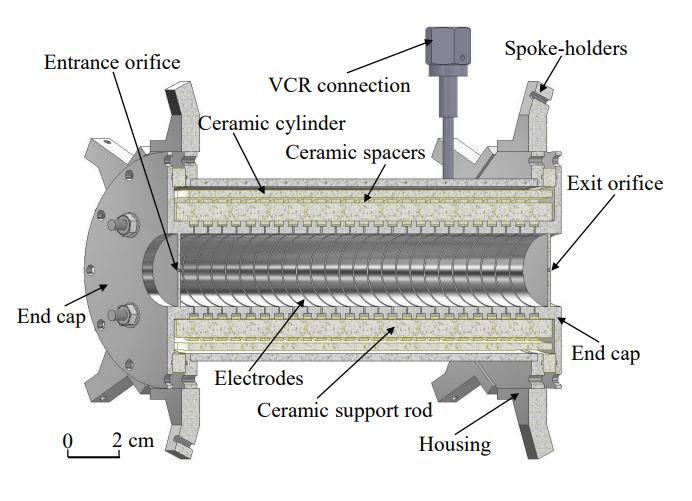}
\caption{Cross-sectional view of the cryogenic drift tube of the IMS apparatus. A resistor chain (not seen in this view) connecting the different drift electrodes is located between these drift electrodes and the drift tube housing.}
\label{fig: ims_drift_cell}
\end{figure} 

The cryogenic drift tube is essentially the heart of the ion mobility spectrometer, where the sample ions are separated by their drift time while traversing the buffer gas medium. 
Figure~\ref {fig: ims_drift_cell} shows the cross-sectional view of this drift tube. 
It is of hexagonal shape with an inner diameter of $46\,$mm and a total length of $153\,$mm. The drift length amounts to $143.5\,$mm, which is given by the distance between the $1.5$ mm orifices of two end caps that axially enclose the drift tube. 
The tube accommodates $24$ stainless steel drift electrodes with an inner diameter of $20\,$mm, an outer diameter of $24\,$mm, and a width of $5\,$mm. A similar type of drift tube (exhibiting a shorter drift length) was developed for the LRC setup and was discussed in detail in Refs.~\cite{Kim:2024, Romero:2022}. 
Although the drift tube is designed to operate between $70$K and $400$K, so far, it has only been operated at $298$K for this work, at which sufficient chromatography performance is expected~\cite{Kim:2024}. 
Helium with a purity of $99.996$\% is used as a buffer gas and is further purified using a Mono-Torr PS4-MT3-R-2 gas getter before injection into the drift tube.

Ions are generated using a laser ablation ion source placed at a distance of $20\,$mm on axis in front of the RF ion buncher as shown in Fig.~\ref{fig: experimental setup}.
It consists of a copper plate to which $0.1$-mm thin, high-purity ($99.9$\%) metal foils of the elements to be examined are attached in such a way that they are aligned at an angle of $\approx 45^{\circ}$ to the laser beam.
In addition, a thin stainless steel plate of $70\,$mm diameter is attached to the source from the rear side for improved ion focusing towards the RF ion buncher.
The ablation laser beam is provided by a Nd:YAG laser (type: Continuum, Precision II) operated in the third harmonic generation mode at a wavelength of $355\,$nm and a repetition rate of $50\,$Hz. 
The laser beam is first shaped with a zoom beam expander (type: EKSMA Optics, \#165-1183) and then focused onto the targeted foil using a UV lens of $500\,$mm focal length.
The position of the ablation beam spot can be scanned using a piezo-driven mirror (type: Thorlabs, KIM101) such that different foils of various elements can be sequentially targeted without breaking the vacuum. 
To prevent the generation of high plasma densities in our measurements, a laser power of less than $20\,$mW in the UV range was completely sufficient for laser ablation. 

For drift time measurements, the ablated ions are first bunched in cycles before injection into the drift tube at a reference time $t_0=0$. For ion bunching, the miniature RF ion buncher is used. It is mechanically fixed to the entrance end cap of the drift tube, similar to the one described in Ref.~\cite{Kim:2024}. 
The only difference is that the buncher used here consists of just $17$ segments and has a total length of approximately $166\,$mm. The last three buncher segments close to the entrance of the drift tube exhibit a similar geometry as given in Ref. \cite{Kim:2024} and are used as a kicker (segment $\rm{S}15$), a trap (segment $\rm{S}16$), and a repeller (segment $\rm{S}17$), respectively.

The miniature ion guide fixed to the exit end cap of the drift tube has a cross-sectional geometry similar to that of the buncher and a total length of $94\,$mm. It consists of $9$ segments, each $10\,$mm long with a separation of 3 mm between the two opposite rods.
Generally, the miniature ion buncher and ion guide operate at RF voltages $\leq35\,$V$_{\text{pp}}$. As these voltages are relatively small, two arbitrary function generators (type: Agilent LXI 33522A) are sufficient to drive the resonance frequency of $990\,$kHz for the corresponding RF circuits.

The other RF ion guide in the intermediate pumping section (PS2) consists of $21$ segments and is $327\,$mm long. Its design is based on the extraction RF quadrupole developed in Ref.~\cite{Neumayr:2006b} and is operated in a similar way. For our experiments, we use an RF amplifier (type: HLA 150) to provide RF voltages $\leq300\,$V$_{\text{pp}}$ required for the operation of this ion guide at a resonance frequency of $1050\,$kHz.
All DC voltages for the different electrodes are provided from a universal multichannel power supply (CAEN, type: SY5527LC) having four 12-channel high voltage boards for positive voltages (type: AG 538DP, $+1.5\,$kV, $10\,$mA) and one 12-channel board for negative voltages (type: AG 538DN, $-1.5\,$kV, $10\,$mA). Channeltron bias voltages are the only ones that are provided separately from a high voltage supply (CAEN, type: N1470, $4\,$CH, $\pm8\,$kV).
Two fast high-voltage transistor switches (type: Behlke HTS 31-06-C) are used to enable a fast switching of the DC voltages for the kicker and the repeller segments of the ion buncher.
We refer the reader to Ref.~\cite{Kim:2024} for a detailed description of the bunching process and the measurement cycle that we typically implement in our measurements. 

\section{Results and discussion}\label{sec:results}
\vspace{-5mm}
\subsection{Characterization of the setup}
\vspace{-5mm}
Laser ablated Lu$^{+}$ is used as a test cation to characterize the spectrometer. 
Figure~\ref{fig:mass_spectra}a shows a corresponding mass spectrum recorded at a QMF mass resolving power of $m/\Delta m \geq 175$. At such a moderate resolving power, the two naturally occurring lutetium isotopes and their monoxides are sufficiently resolved, and we have chosen the $^{175}$Lu$^{+}$ isotope for further measurements.
\begin{figure}[t!]
\centering
\includegraphics[width=0.48\textwidth] {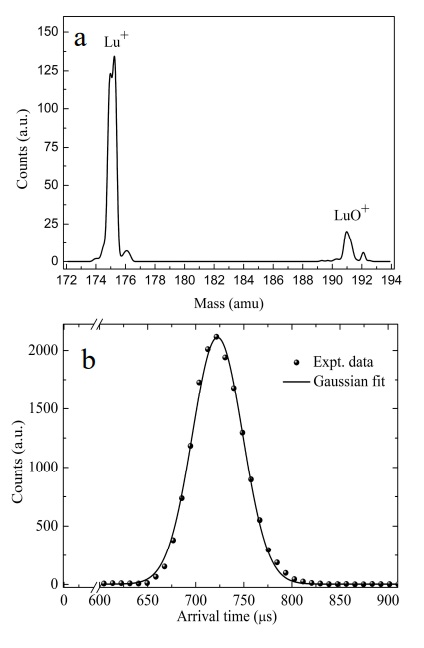}
\caption{\textbf{a}: Mass spectrum recorded during laser ablation from a lutetium metallic foil. \textbf{b}: Arrival time distribution of $^{175}$Lu$^{+}$ cations drifting in He at a pressure of $1.5\,$mbar and at a reduced electric field of $8\,$Td. Only a moderate power of $5\,$mW was used for laser ablation. Black solid line: Best Gaussian fit to the data.}
\label{fig:mass_spectra}
\end{figure} 
To evaluate the performance of the drift tube, we first studied the mean arrival times of Lu$^+$ ions at helium pressures ranging from $1.5$ to $8\,$mbar and reduced electric fields varying from $1$ to $20\,$Td. 
Figure~\ref{fig:mass_spectra}b shows the arrival time distribution (ATD) recorded at $1.5\,$mbar and $8\,$Td. The mean arrival time ($t_{\text{ATD}}$) is extracted from the best fit to the measured data, whereas in most cases a simple Gaussian distribution is sufficient to mimic the recorded arrival time distributions. The peak width in terms of full width at half maximum (FWHM) is obtained from the fit accordingly. 
In our experiments, the $t_{\text{ATD}}$ values do not change significantly at a fixed reduced field, given the buffer gas pressure is $\geq 2\,$mbar, which implies a nearly stationary condition for ion drift at higher pressures.
As the length of the drift tube is short ($\approx15\,$cm), increasing the drift pressure also helps to increase the resolution of the spectrometer~\cite{Tabrizchi:2006} and thus promises more accurate IMS measurements. 
In the presented data, the mean arrival time of the ions consists of two components: the drift time $t_{\text{d}}$, which is the time the ions need on average to traverse the drift tube and the offset time of flight $t_{\text{offset}}$, which they spent outside of the tube on their way to the detector. At a fixed buffer gas pressure, both $t_{\text{ATD}}$ and $t_{\text{d}}$ vary inversely proportional to the electric field strength. 
Figure~\ref{fig: time resolution of IMS set up} shows the relative time width of the ions, i.e. the FWHM to the mean arrival time, for different He pressures as a function of the reduced electric field. 
In general, the relative width (FWHM/$t_{\text{ATD}}$) scales with $\sqrt{T/E}$ for narrowest initial pulse width~\cite{Rokushika:1985}. 
As can be seen in Fig.~\ref{fig: time resolution of IMS set up}, the relative widths decrease rapidly with increasing reduced fields, following the expected trend. 
At higher electric fields, however, the relative width slowly increases again because the drift time decreases, while the FWHM remains nearly constant. The stagnation of the FWHM at higher fields arises from the finite time width of the initial ion bunch and the combined effect of longitudinal diffusion inside and outside of the drift tube.

In addition, we measure a rapid decrease in the relative width with increasing gas pressure in the drift tube, as one would expect in the case of a non-negligible initial width of the ion bunch~\cite{Tabrizchi:2006}.
In our experiments, determining the initial width is challenging, as it is influenced by the background pressure in the chromatography section (PS1) between the trap section and the entrance of the drift tube, as well as by the cooling time during the bunching process itself. 
Moreover, we observed that for a fixed reduced field, ion transmission can be substantially improved by operating the drift tube at higher gas pressures. This is because frequent collisions at higher pressures result in reduced collisional mean free paths, thereby enhancing ion confinement near the central axis of the drift tube. 
Following optimization of the IMS operation, we achieved a minimum relative time width of $6$\% at $8\,$mbar pressure and $4\,$Td reduced field, which should allow us to distinguish between the distinct and state-specific ion mobilities of Lu$^{+}$ ions drifting in helium at room temperature~\cite{Laatiaoui:2020b,Kim:2024}.\\ 
\begin{figure}[thb!]
\centering
\includegraphics[width=0.48\textwidth] {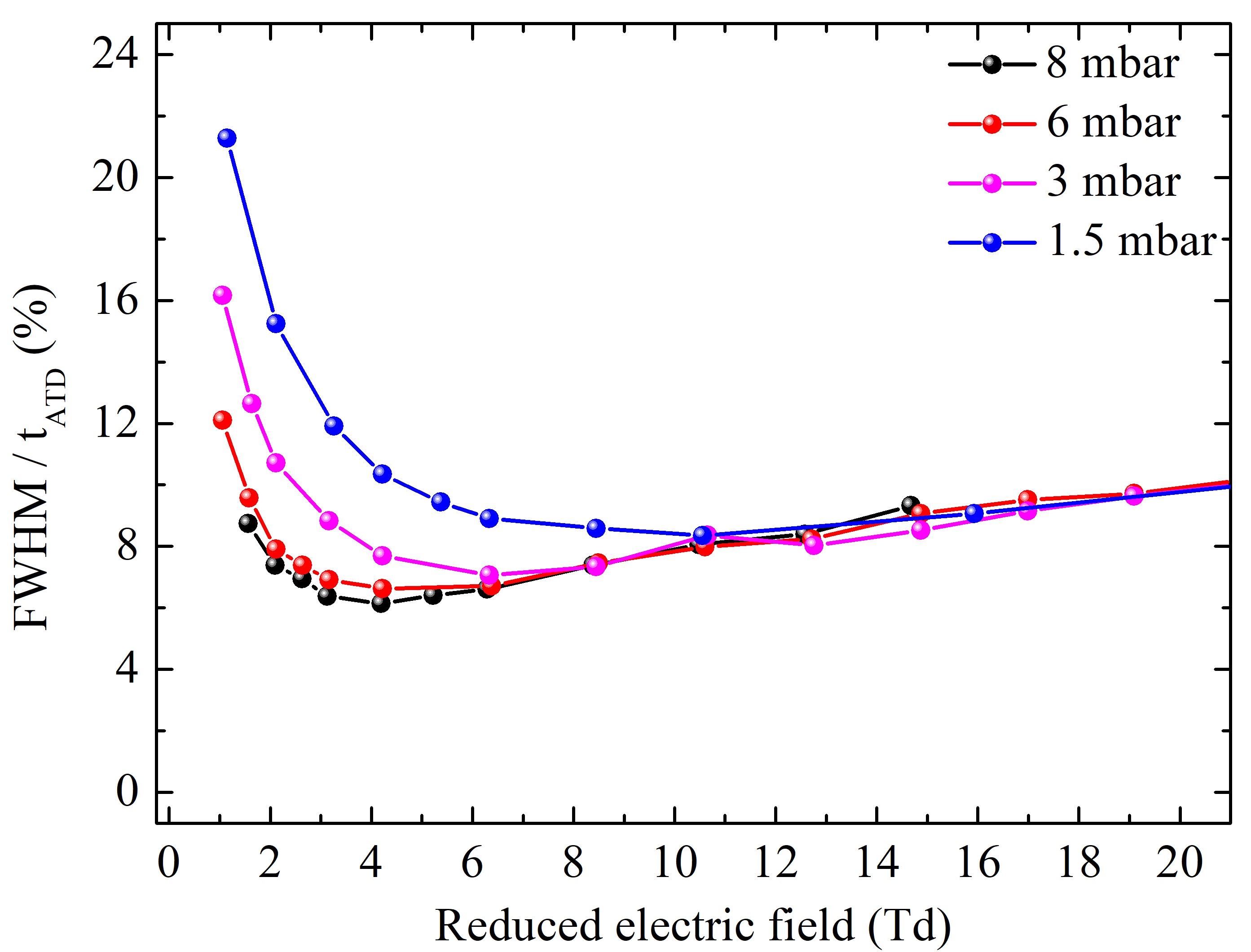}
\caption{Relative time widths at different He pressures for Lu$^+$ ions in the ground state as a function of the reduced electric field.}
\label{fig: time resolution of IMS set up}
\end{figure}
\begin{figure}[bht!]
\centering
\includegraphics[width=0.48\textwidth] {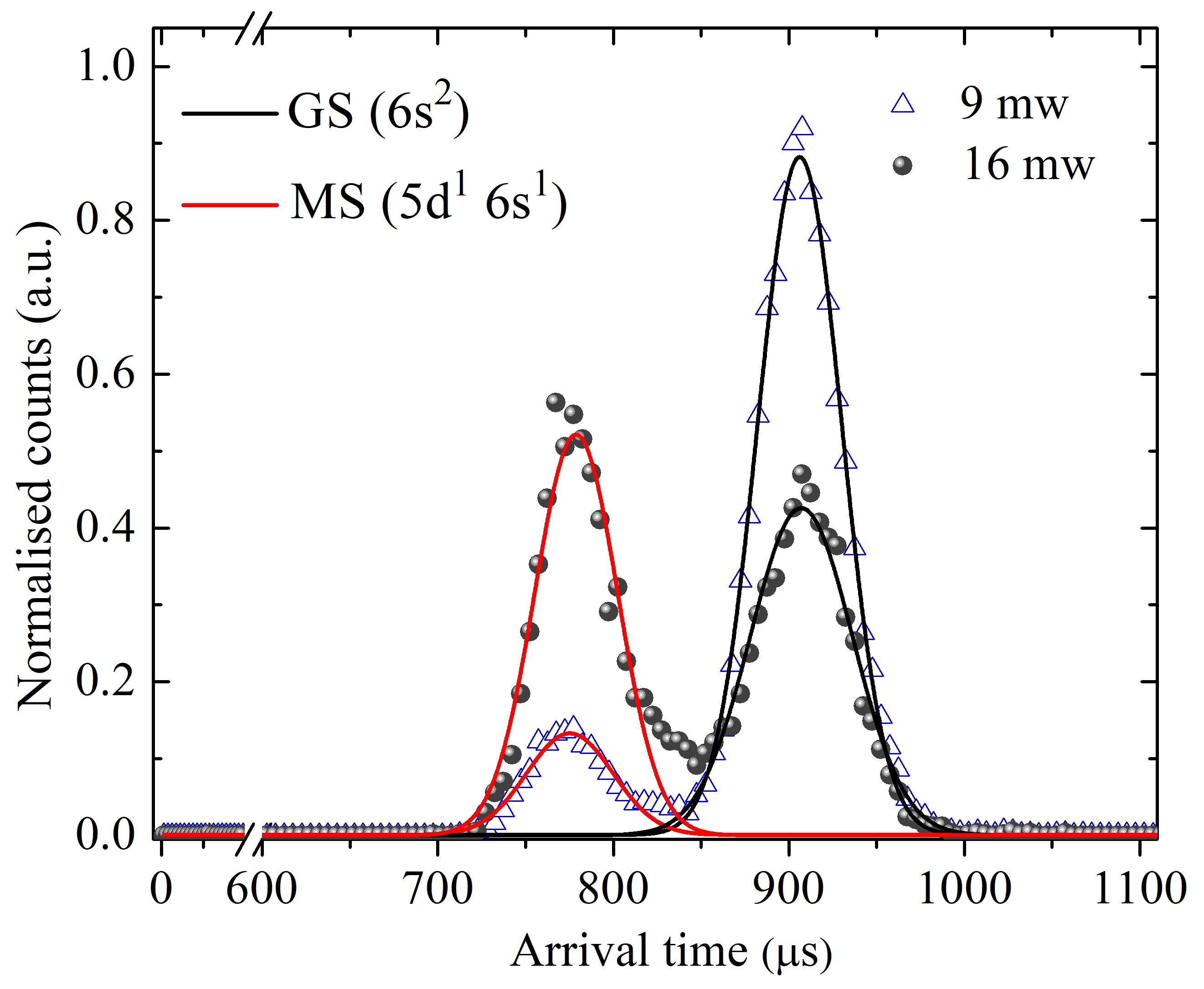}
\caption{Measured arrival time distributions of Lu$^+$ ions that were ablated at $9\,$mW (\textcolor{blue}{$\triangle$}) and $16\,$mW (\textcolor{darkgray} {$\bullet$}) laser powers, at a reduced field of $5\,$Td, and at a drift tube pressure of $8\,$mbar. The solid lines indicate Gaussian best fits to the data. See text for more details.}
\label{fig: ESC_Lu}
\end{figure}

\subsection{Extraction of the state-specific ion mobilities}
\vspace{-5mm}
The Electronic State Chromatography effect in Lu$^+$ has already been observed during the inauguration experiments of the laser resonance chromatography apparatus described in Ref.~\cite{Kim:2024}. However, an accurate value for the metastable-state ion mobility of this cation drifting in He could not be determined yet due to the very short drift tube that was used in these former experiments. In the present work, we observed the ESC effect in Lu$^+$ as well, utilizing a drift tube that is approximately three times longer than that used in Ref.~\cite{Kim:2024}. Figure~\ref{fig: ESC_Lu} shows the measured arrival time distributions of Lu$^+$ at $8\,$mbar He pressure and $5\,$Td reduced field for different laser power for ablation of $9\,$mW and $16\,$mW. In each of the recorded time spectra, two distinct peaks can be resolved and identified. 
The right peaks correspond to the arrival time of the slower ground-state (GS) ions, whereas the left peaks correspond to the faster metastable state (MS) ions. These peak assignments are rather straightforward. As shown in Fig.~\ref{fig: ESC_Lu}, the fraction of faster ions increases with increasing laser power for ablation at the expense of a decrease in the fraction of slower ions. 
From these measurements, we infer that Lu$^+$ in the ground state ($6s^2$) with two $6s$ valence electrons interacts more strongly with helium atoms than it does in the lowest metastable states, which possess just one $6s$ valence electron ($5d^16s^1$), and thus is lagging behind.
In the intermediate region between the distinct peaks, we observed some events that bridge the two peaks in the arrival time distributions. We attribute them to state deactivation processes in which the population of the metastable state is quenched to the ground state by gas collisions during ion drift. We therefore excluded data from this intermediate region during the peak-fitting procedures to extract the state-specific mean arrival times. 
Since the left peaks have similar relative widths as the right ones, we assign the left peaks to only one dominant state, namely the lowest-lying metastable $^3$D$_1$ state in Lu$^+$. This assignment is fully justified since gas collisions promote intra-multiplet quenching, leading to a strong feeding into the $^3$D$_1$ state from the energetically closer upper states.
Unlike in Ref.~\cite{Kim:2024}, we did not observe a shift of the ground state peak to shorter times due to the population of higher-lying metastable states, as we have applied only moderate laser powers in our experiments so far. 

Since the mean arrival time is composed of a drift time and an offset time, it can be expressed as \cite{Manard:2017}:
\begin{equation}\label{Eq1}
t_{\text{ATD}} = t_{\text{d}} + t_{\text{offset}} = \left(\frac{L^2 T_0}{K_0 P_0 T}\right) \left(\frac{P}{V}\right) + t_{\text{offset}}.
\end{equation}
Here, $V$ is the voltage difference across the length $L$ of the drift tube. $P$ and $T$ are gas pressure and temperature. 
In Fig.~\ref{fig:LinearFittings} we show mean arrival times obtained for different ratios of gas pressure to voltage difference ($P/V$). 
Since the drift time scales linearly with this ratio, the reduced ion mobilities for the different states can be extracted directly from the slopes of linear fits to the data according to Eq.~\ref{Eq1}. 
In general, the measured reduced mobilities have an uncertainty of $2$\%, arising mainly from buffer gas pressure fluctuations of approximately $\pm0.1\,$mbar. The resulting low-field reduced mobilities of Lu$^+$ are thus ($16.6\pm0.2$)~cm$^2$/Vs and ($19.7\pm0.3$)~cm$^2$/Vs for the ground and metastable states, respectively. 
These are the average mobility values obtained from five consecutive measurements of the aforementioned slopes. All measurements were conducted at reduced fields of up to $5\,$Td and gas pressures of at least $2\,$mbar, at which homogeneous ion drift at low fields can be safely assumed.

Table~\ref{tab:Values} shows an overview of measured ion mobilities and the corresponding diffusion coefficients along with literature values. The state-specific diffusion coefficient (given in terms of n$_{0}$D) is obtained from the corresponding low-field mobility in accordance with the Einstein relation~\cite{Mason:1988}, which holds in the low-field regime as in our case.
The measured reduced ion mobility for the ground state agrees well with the value reported by Manard and Kemper~\cite{Manard:2017} from their high-precision IMS experiments. In addition, it is in excellent agreement with the theoretically predicted values based on \textit{ab-initio} interaction potentials~\cite{Visentin:2020,Laatiaoui:2020b,Ramanantoanina:2023}. Our measured value for Lu$^+$ in the metastable state is a first-time addition to the literature. It agrees with the value of $19.5\,$cm$^2$/Vs obtained based on interaction potentials from Multi-Reference Configuration Interaction (MRCI)~\cite{Ramanantoanina:2023} calculations. Predictions based on interaction potentials from Scalar-Relativistic and Spin-Orbit coupling (SR+SO) calculations~\cite{Laatiaoui:2020b}, though consistent with our data, they tend to slightly overestimate the low-field mobility of the metastable state. We trace back this to the degree of accuracy applied to determine the isotropic interaction potential for the metastable state $V_0(^3\text{D}_1)$~\cite{Laatiaoui:2020b}. 
The relative difference in ion mobility between the ground- and metastable state is found to be $15.7$\%. 
\begin{figure}[t!]
\centering
\includegraphics[width=0.48\textwidth] {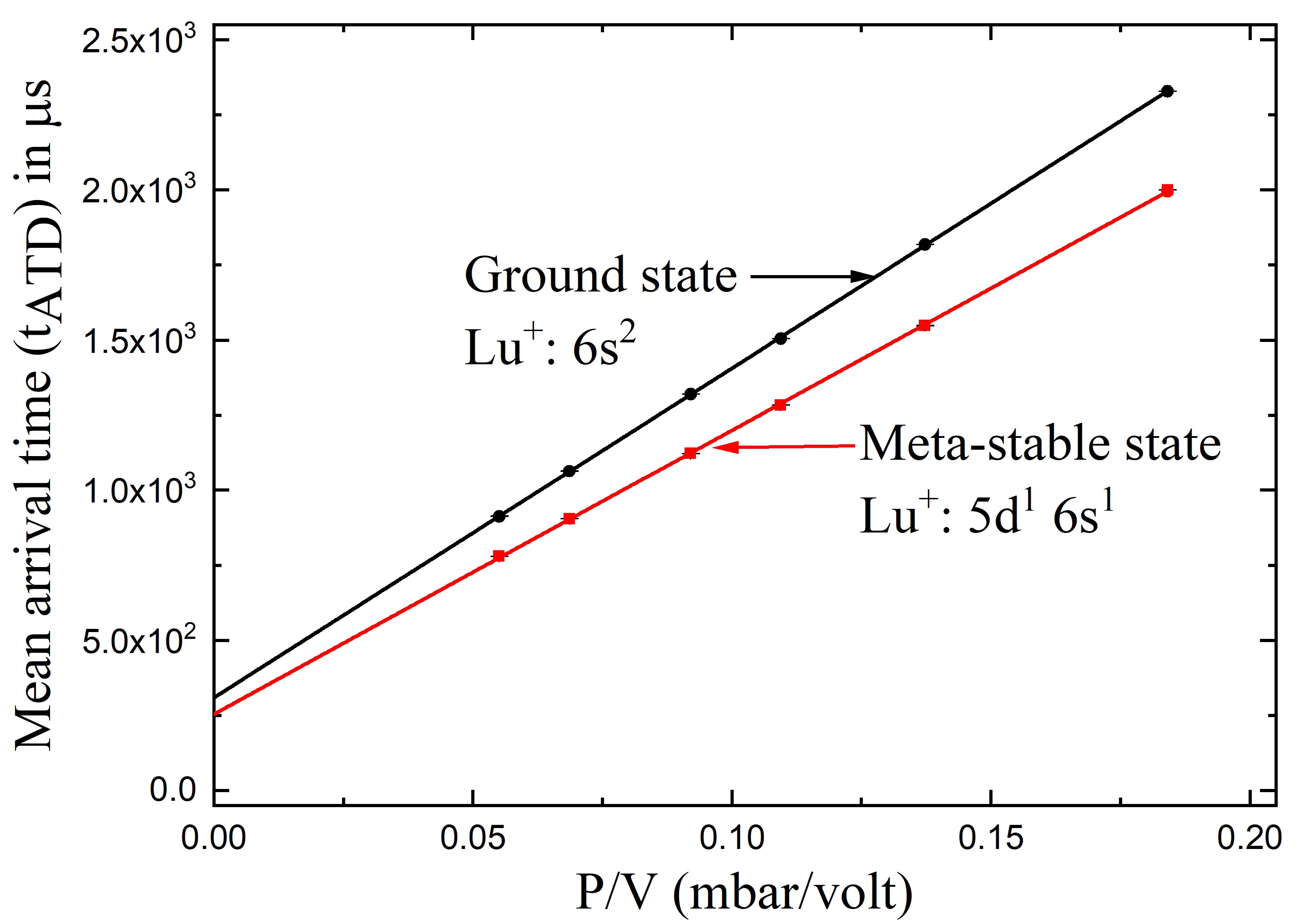}
\caption{Arrival time of Lu$^+$ ions for ground- ($\bullet$) and metastable (\textcolor{red}{${_\blacksquare}$}) states as function of $P/V$. Solid lines are the best fits to the corresponding data sets according to Eq.~\ref{Eq1}.}
\label{fig:LinearFittings}
\end{figure}
\begin{table}[tb]
   \caption{Measured (This work) ion mobility $K_0$ and diffusion coefficient $n_0D$ of Lu$^+$ in He drifting in the ground- (GS) and metastable state (MS) in comparison with reported values from theory based on SR+SO ($\ast$) and MRCI ($\star$) interaction potentials as well as a single mobility value from experiment ($^\#$).}

    \centering
    \begin {tabular}{lcccccccccc}    
         \hline
         \hline
         Property                                 && State && This work && $\ast$\cite{Laatiaoui:2020b} && $\star$\cite{Ramanantoanina:2023} && $^\#$\cite{Manard:2017}\\
         \hline
         \multirow{2}{*}{$K_0$ (cm$^2$/V s)}      && GS    && 16.6(3)   && 16.6                         && 16.5                              && 16.80(4)               \\
                                                  && MS    && 19.7(4)   && 20.6                         && 19.5                              && --                     \\
         \hline 
         \multirow{2}{*}{$n_0D$ (10$^{19}$/cm s)} && GS    && 1.14(2)   && 1.15                         && --                              && --                     \\
                                                  && MS    && 1.35(3)   && 1.41                         && --                              && --                     \\
         \hline
         \hline
    \end{tabular}
    \label{tab:Values}
\end{table}

Figure~\ref{fig:IonMobilityWithField} shows the variation of reduced mobility with the reduced field up to $20\,$Td, along with theoretically estimated values based on SR+SO~\cite{Laatiaoui:2020b} and MRCI~\cite{Ramanantoanina:2023} interaction potentials. Generally there is a good agreement between theory and experiment. The reduced ion mobility of Lu$^+$ remains nearly constant at $E/n_0<10\,$Td, as expected, and starts to decrease gradually with increasing reduced fields. 
According to theory, the kinetic energy of the ion increases with increasing fields and leads to an increase in the effective ion temperature beyond that of the surrounding buffer gas.
Collisions begin to probe the repulsive wall of the interaction potential at relatively smaller internuclear distances. Hence, the interactions last longer and the collision cross section increases, thereby decreasing the ions' reduced mobility. 
For accurate IMS studies at high reduced fields beyond 20 Td, it is however necessary that the ions of interest reach their terminal velocity quickly when injected into the drift tube in order to minimize systematic uncertainties. This can be achieved either by utilizing elongated drift tubes or by operating the drift tube at elevated buffer gas pressures. Since our cryogenic drift tube is designed to be of intermediate length to enable electronic state chromatography studies, we were able to measure Lu$^+$ mobilities up to a reduced field of $20\,$ Td. This was limited, on the one hand, by the maximum electric field that can be applied at high pressures before discharges occur and, on the other hand, by the non-ideal condition of non-homogeneous ion drift during ion injection at low pressures~\cite{Kim:2024}. 
Moreover, we have restricted our systematic studies to low $E/n_0$ values, where we can still measure ATDs of Lu$^+$ in the metastable states with sufficient statistics despite collisional de-excitations.
\begin{figure}[tb]
\centering
\includegraphics[width=0.48\textwidth] {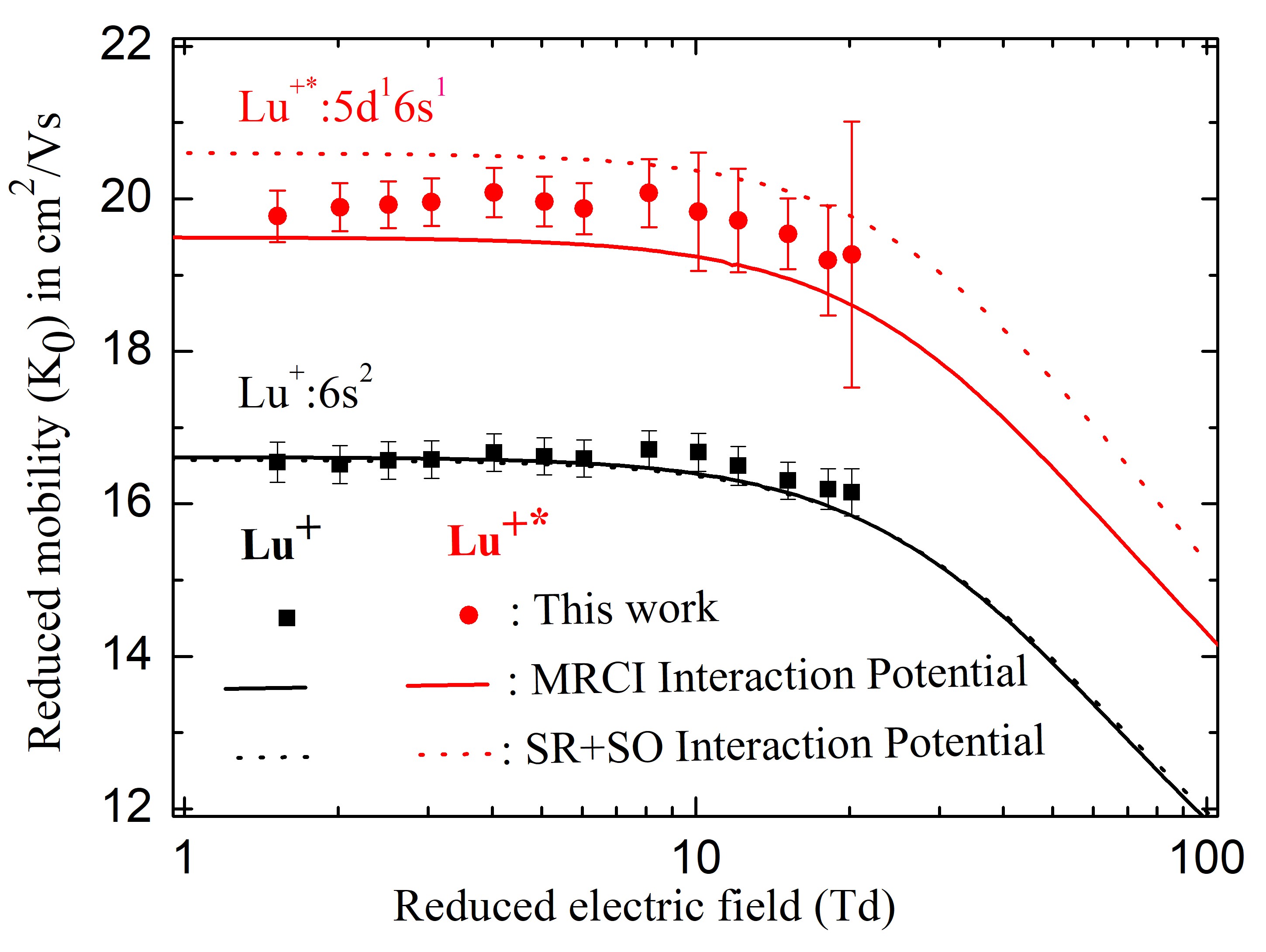}

\caption{Comparison of measured reduced mobilities of both lutetium ground (Lu$^+$) and metastable states (Lu$^{+*}$) ions with theoretical predictions as a function of the reduced electric field. Measured data points: Lu$^+$($_\blacksquare$) and Lu$^{+*}$(\textcolor{red}{$\bullet$}). Solid lines represent mobility predictions based on MRCI~\cite{Ramanantoanina:2023}, Lu$^+$(\textcolor{black}{---}) and Lu$^{*+}$(\textcolor{red}{---}) interaction potentials while dotted lines correspond to same using SR+SO ~\cite{Laatiaoui:2020b}, Lu$^+$(\textcolor{black}{...}) and Lu$^{*+}$(\textcolor{red}{...}).}

\label{fig:IonMobilityWithField}
\end{figure}

\section{Summary and outlook}\label{sec:summary}
\vspace{-5mm}
A new cryogenic ion mobility spectrometer for heavy element research has been developed and characterized. 
Electronic state chromatography for Lu$^+$ ions has been demonstrated by measuring their low-field reduced mobilities in helium at $298$K in both ground and metastable states. The measured values show good agreement with the values in the literature, including an experimental value from high-precision IMS measurements. The mobility deviation between the ion's states amounts to $15.7$\% as expected from the theory. Additionally, the variation of the reduced mobility of Lu$^+$ in both states with reduced electric field has been examined.
In the next step, we envisage to study the low-field mobilities at cryogenic temperatures, thereby assessing the chromatographic performance of our setup, before extending such measurements to the actinides. Since ion mobility is sensitive to electronic configurations, we hope that this research will deepen our understanding of the interplay between transport properties and relativistic effects in the region of the heaviest elements. 

\section*{Declaration of competing interest}
\vspace{-5mm}
The authors declare that they have no known competing financial interests or personal relationships that could have appeared to influence the work reported in this paper.

\section*{Data availability}
\vspace{-5mm}
Data will be made available on request. Transport coefficients and interaction potentials resulting from our experiments will be made available for public use within the framework of the Plasma Data Exchange Project LXcat. A corresponding open-access website can be found under: https://nl.lxcat.net/home/. 

\section*{Acknowledgements}
\vspace{-5mm}
This project has received funding from the Deutsche Forschungs Gemeinschaft (DFG), German Research Foundation -- Project No. 426500921 and the European Research Council (ERC) under the European Union’s Horizon 2020 Research and Innovation Programme -- Grant Agreement No. 819957.


{\addtolength{\textheight}{1\baselineskip}

\bibliography{main}

@article{Giorgio:2024,
  title={Transport-property predictions for laser resonance chromatography on Rf+(Z= 104)},
  author={Visentin, Giorgio and Ramanantoanina, Harry and Borschevsky, Anastasia and Viehland, Larry and Jana, Biswajit and Arya, Aayush and Fritzsche, Stephan and Laatiaoui, Mustapha},
  journal={Physical Review A},
  volume={110},
  number={1},
  pages={012805},
  year={2024},
  publisher={APS}
}

@article{Backe:1997,
title = {A compact apparatus for mass selective resonance ionization spectroscopy in a buffer gas cell},
author = {H. Backe and K. Eberhardt and R. Feldmann and M. Hies and H. Kunz and W. Lauth and R. Martin and H. Schöpe and P. Schwamb and M. Sewtz and P. Thörle and N. Trautmann and S. Zauner},
journal = {Nuclear Instruments and Methods in Physics Research Section B: Beam Interactions with Materials and Atoms},
volume = {126},
number = {1},
pages = {406-410},
year = {1997},
doi = {https://doi.org/10.1016/S0168-583X(96)01035-X},
}

@article{Rokushika:1985,
  title={Resolution Measurement for Ion Mobility Spectrometry},
  author={Rokushika, S. and Hatano, H. and Baim, M. and Hill, H.},
  journal={Anal. Chem.},
  volume={57},
  pages={1902},
  year={1985},
}

@article{Tabrizchi:2006,
  title={Pressure effects on resolution in ion mobility spectrometry},
  author={Tabrizchi, M. and Rouholahnejad, F.},
  journal={Talanta},
  volume={69},
  pages={87},
  year={2006},
}

@article{Kim:2024,
      title={Laser Resonance Chromatography: First Commissioning Results and Future Prospects}, 
      author={Kim, EunKang and Jana, Biswajit and Arya, Aayush and Block, Michael and Raeder, Sebastian and Ramanantoanina, Harry and Rickert, Elisabeth and Romero Romero, Elisa and Laatiaoui, Mustapha},
      journal   = {Nuclear Instruments and Methods in Physics Research Section B},
      volume={555},
      year={2024},
      pages={165461},
      artnum={165461},
}

@Article{Romero:2022,
AUTHOR = {Romero Romero, Elisa and Block, Michael and Jana, Biswajit and Kim, Eunkang and Nothhelfer, Steven and Raeder, Sebastian and Ramanantoanina, Harry and Rickert, Elisabeth and Schneider, Jonas and Sikora, Philipp and Laatiaoui, Mustapha},
TITLE = {A Progress Report on Laser Resonance Chromatography},
JOURNAL = {Atoms},
VOLUME = {10},
YEAR = {2022},
NUMBER = {3},
ARTICLE-NUMBER = {87},
PAGES = {87},
}

@incollection{Valerie:2021,
    author = {Gabelica, Valérie},
    isbn = {978-1-83916-166-7},
    title = "{Ion Mobility–Mass Spectrometry: an Overview}",
    booktitle = "{Ion Mobility – Mass Spectrometry: Fundamentals and Applications}",
    publisher = {The Royal Society of Chemistry},
    year = {2021},
    month = {11},
    doi = {10.1039/9781839162886-00001},
    url = {https://doi.org/10.1039/9781839162886-00001},
}

@article{Wybourne:2002,
  title={Relativistic effects in lanthanides and actinides},
  author={Wybourne, Brian G and Smentek, Lidia},
  journal={Journal of alloys and compounds},
  volume={341},
  number={1-2},
  pages={71--75},
  year={2002},
  publisher={Elsevier}
}

@article{Pyykko:2012,
  title={Relativistic effects in chemistry: more common than you thought},
  author={Pyykk{\"o}, Pekka},
  journal={Annual review of physical chemistry},
  volume={63},
  pages={45--64},
  year={2012},
  publisher={Annual Reviews}
}

@article{Schadel:2015,
  title={Chemistry of the superheavy elements},
  author={Sch{\"a}del, Matthias},
  journal={Philosophical Transactions of the Royal Society A: Mathematical, Physical and Engineering Sciences},
  volume={373},
  number={2037},
  pages={20140191},
  year={2015},
  publisher={The Royal Society Publishing}
}

@book{Wilkins:2010,
  title={Ion mobility spectrometry-mass spectrometry: theory and applications},
  author={Wilkins, Charles L and Trimpin, Sarah},
  year={2010},
  publisher={CRC press},
  ISBN={978-1439813249}
}

@article{Visentin:2020,
  title={Mobility of the singly-charged lanthanide and actinide cations: Trends and perspectives},
  author={Visentin, Giorgio and Laatiaoui, Mustapha and Viehland, Larry A and Buchachenko, Alexei A},
  journal={Frontiers in Chemistry},
  volume={8},
  pages={438},
  year={2020},
  publisher={Frontiers Media SA}
}

@article{Schwerdtfeger:2015,
  title={Relativistic and quantum electrodynamic effects in superheavy elements},
  author={Schwerdtfeger, Peter and Pa{\v{s}}teka, Luk{\'a}{\v{s}} F and Punnett, Andrew and Bowman, Patrick O},
  journal={Nuclear Physics A},
  volume={944},
  pages={551--577},
  year={2015},
  publisher={Elsevier}
}

@article{Tatewaki:2017,
  title={Relativistic effects in the electronic structure of atoms},
  author={Tatewaki, Hiroshi and Yamamoto, Shigeyoshi and Hatano, Yasuyo},
  journal={ACS omega},
  volume={2},
  number={9},
  pages={6072--6080},
  year={2017},
  publisher={ACS Publications}
}

@article{Das:2023,
  title={Relativistic effects on the chemistry of heavier elements: why not given proper importance in chemistry education at the undergraduate and postgraduate level?},
  author={Das, Ankita and Das, Udita and Das, Ruhi and Das, Asim K},
  journal={Chemistry Teacher International},
  volume={5},
  number={4},
  pages={365--378},
  year={2023},
  publisher={De Gruyter}
}

@Article{Laatiaoui:2012,
  author    = {Laatiaoui, M. and Backe, H. and Habs, D. and Kunz, P. and Lauth, W. and Sewtz, M.},
  journal   = {The European Physical Journal D},
  title     = {Low-field mobilities of rare-earth metals},
  year      = {2012},
  pages     = {232},
  volume    = {66},
  doi       = {10.1140/epjd/e2012-30221-3},
}

@article{Sewtz:2003,
  title = {First Observation of Atomic Levels for the Element Fermium ($Z=100$)},
  author = {Sewtz, M. and Backe, H. and Dretzke, A. and Kube, G. and Lauth, W. and Schwamb, P. and Eberhardt, K. and Gr\"uning, C. and Th\"orle, P. and Trautmann, N. and Kunz, P. and Lassen, J. and Passler, G. and Dong, C. Z. and Fritzsche, S. and Haire, R. G.},
  journal = {Phys. Rev. Lett.},
  volume = {90},
  issue = {16},
  pages = {163002},
  numpages = {4},
  year = {2003},
  month = {Apr},
  publisher = {American Physical Society},
  }

@Article{Block:2021,
  author    = {Michael Block and Mustapha Laatiaoui and Sebastian Raeder},
  journal   = {Progress in Particle and Nuclear Physics},
  title     = {Recent progress in laser spectroscopy of the actinides},
  year      = {2021},
  month     = {jan},
  pages     = {103834},
  volume    = {116},
  doi       = {10.1016/j.ppnp.2020.103834},
  publisher = {Elsevier {BV}},
}

@Article{Rickert:2020,
  title={Ion mobilities for heaviest element identification},
  author={Rickert, Elisabeth and Backe, Hartmut and Block, Michael and Laatiaoui, Mustapha and Lauth, Werner and Raeder, Sebastian and Schneider, Jonas and Schneider, Fabian},
  journal={Hyperfine Interactions},
  volume={241},
  number={1},
  pages={49},
  year={2020},
  publisher={Springer}
}

@Article{Laatiaoui:2020a,
  author    = {Mustapha Laatiaoui and Alexei A. Buchachenko and Larry A. Viehland},
  journal   = {Physical Review Letters},
  title     = {Laser Resonance Chromatography of Superheavy Elements},
  year      = {2020},
  month     = {jul},
  number    = {2},
  pages     = {023002},
  volume    = {125},
  doi       = {10.1103/physrevlett.125.023002},
  publisher = {American Physical Society ({APS})},
}

@Article{Laatiaoui:2020b,
  author    = {Mustapha Laatiaoui and Alexei A. Buchachenko and Larry A. Viehland},
  journal   = {Physical Review A},
  title     = {Exploiting transport properties for the detection of optical pumping in heavy ions},
  year      = {2020},
  month     = {jul},
  number    = {1},
  pages     = {013106},
  volume    = {102},
  doi       = {10.1103/physreva.102.013106},
  publisher = {American Physical Society ({APS})},
}

@Article{Neumayr:2006b,
  author    = {Neumayr, JB. and Thirolf, PG and Habs, D. and Heinz, S. and Kolhinen, V. S. and Sewtz, M. and Szerypo, J.},
  journal   = {Rev. Sci. Instrum.},
  title     = {{Performance of the MLL-IonCatcher}},
  year      = {2006},
  number    = {6},
  pages     = {065109},
  volume    = {77},
  publisher = {AIP Publishing},
}

@Article{Backe:2015,
  author    = {Backe, H. and Lauth, W. and Block, M. and Laatiaoui, M.},
  journal   = {Nuclear Physics A},
  title     = {Prospects for laser spectroscopy, ion chemistry and mobility measurements of superheavy elements in buffer-gas traps},
  year      = {2015},
  pages     = {492--517},
  volume    = {944},
  publisher = {Elsevier},
}

@Article{Sewtz:2003b,
  author    = {Sewtz, M. and Backe, H. and Dong, CZ and Dretzke, A and Eberhardt, K and Fritzsche, S and Gr{\"u}ning, C and Haire, RG and Kube, G and Kunz, P and others},
  journal   = {Spectrochimica Acta Part B: Atomic Spectroscopy},
  title     = {Resonance ionization spectroscopy of fermium (Z= 100)},
  year      = {2003},
  number    = {6},
  pages     = {1077--1082},
  volume    = {58},
  publisher = {Elsevier},
}

@Article{Kemper:1991,
  author  = {Kemper, P.R. and Bowers, M.T.},
  title   = {{E}lectronic-state chromatography: application to first-row transition-metal ions},
  journal = {J. Phys. Chem.},
  year    = {1991},
  volume  = {95},
  pages   = {5134-5146},
}

@Article{Laatiaoui:2019,
  author  = {Laatiaoui, M. and Raeder, S.},
  title   = {{L}aser {S}pectroscopy of the {H}eaviest {E}lements: {O}ne {A}tom at a {T}ime},
  journal = {Nucl. Phys. News},
  year    = {2019},
  volume  = {29},
  number  = {1},
  pages   = {21-25},
  doi     = {https://doi.org/10.1080/10619127.2019.1571804},
}

@Book{Mason:1988,
  title     = {{T}ransport {P}roperties of {I}ons in {G}ases},
  publisher = {John Wiley and Sons},
  year      = {1988},
  author    = {Mason, E. and McDaniel, E.},
  address   = {New York},
  ISBN={9783527602858},
}

@Article{Manard:2017,
  author  = {Manard, M.J. and Kemper, P.R.},
  title   = {{R}educed mobilities of lanthanide cations measured usinghigh-resolution ion mobility mass spectrometry with comparisons between experiment and theory},
  journal = {Int. J. Mass Spectrom.},
  year    = {2017},
  volume  = {412},
  pages   = {14},
}

@ARTICLE{Manard:2017b,
author= "Manard, M. J. and Kemper, P. R.",
title=  "{A}n experimental investigation into the reduced mobilities of lanthanide cations using high-resolution ion mobility mass spectrometry",
journal="International Journal of Mass Spectrometry",
volume= "423",
number= "",
pages=  "54-58",
year=   "2017",
doi=    "10.1016/j.ijms.2017.10.010",
    }

@inbook{Armentrout:2011,
    author = "Armentrout, P. B.",
    title ="{E}lectronic {S}tate {C}hromatography in Ion Mobility Spectrometry-Mass Spectrometry" ,
    publisher="CRC Press, Taylor \& Francis Group",
    year = "2010",
    chapter = "2.4",
    ISBN ="978-1439813249",
}

@ARTICLE{Bowers:1993,
author= "Bowers, M. T. and Kemper, P. R. and von Helden, G. and van Koppen, P. A. M.",
title=  "{G}as-{P}hase {I}on {C}hromatography: {T}ransition {M}etal {S}tate {S}election and {C}arbon {C}luster {F}ormation",
journal="Science",
volume= "260",
pages=  "1446",
year=   "1993",
doi=    "",
    }

@Article{Ramanantoanina:2023,
  author    = {Ramanantoanina, H. and others},
  journal   = {Physical Review A},
  title     = {{S}tate-specific ion mobilities of {L}r$^+$ ({$Z=103$}) in helium},
  year      = {2023},
  month     = {},
  number    = {},
  volume    = {108},
  pages     = {012802},
  doi       = {10.1103/PhysRevA.108.012802},
  publisher = {American Physical Society ({APS})},
}
\end{document}